\documentclass[twocolumn,showpacs,preprintnumbers,amsmath,amssymb]{revtex4}
\usepackage{txfonts}
\usepackage{mathrsfs}
\usepackage{amssymb}

\usepackage{graphicx}
\usepackage{bm}

\begin{document}
\title{Quantum repeaters free of polarization disturbance and phase noise}
\author{Zhen-Qiang Yin, Yi-Bo Zhao, Yong Yang, Zheng-Fu Han*, Guang-Can Guo}
\affiliation{Key Laboratory of Quantum Information\\ University of
Science and Technology of China\\ Hefei 230026\\ China}

\begin{abstract}
 Original quantum repeater protocols based on
single-photon interference suffer from phase noise of the channel,
which makes the long-distance quantum communication infeasible.
Fortunately, two-photon interference type quantum repeaters can be
immune to phase noise of the channel. However, this type quantum
repeaters may still suffer from polarization disturbance of the
channel. Here we propose a quantum repeaters protocol which is free
of polarization disturbance of the channel based on the invariance
of the anti-symmetric Bell state
$|\psi^-\rangle=(|H\rangle|V\rangle-|V\rangle|H\rangle)/\sqrt{2}$
under collective noise. Our protocol is also immune to phase noise
with the Sagnac interferometer configuration. Through single-atom
cavity-QED technology and linear optics, this scheme can be
implemented easily.
\end{abstract}


\maketitle

\section{introduction}
 Many applications of quantum communication, such as quantum
cryptography \cite{BB84, Ekert1991}, rely on distributing quantum
states between two distant peers, called Alice and Bob. However, the
real-life channel will inevitably absorb or add noises to the
information carriers (flying photons), which makes the efficiency of
distributing quantum states decrease exponentially  with the length
of channel. To implement long-distance ($\geqslant 1000km$) quantum
communication, this efficiency should scale polynomially with the
length of channel. With the help of quantum repeaters \cite{quantum
repeaters}, one can implement long distance quantum communication in
principle. The basic ideas of quantum repeaters are: 1) dividing the
transmission channel between Alice and Bob into many short segments
and distributing entangled-states between the neighboring nodes,
each of which can be regarded as individual quantum-memory units, 2)
through entanglement swapping and purification \cite{entanglement
swapping1, entanglement swapping2}, the range of entanglement could
be extended, 3) finally entangled-states are distributed between
Alice and Bob.

 The physical implementations of quantum repeaters can be based on
single atom trapped in cavity \cite{QR4, QR5} or atomic ensembles
with linear optics (DLCZ) \cite{QR3}. For the experimental
progresses, one can see Refs. \cite{QRE1, QRE2, QRE3, QRE4, QRE5,
QRE6, QRE7}. Most of quantum repeaters protocols utilize the same
topological structure as DLCZ's. Suppose we have two neighboring but
distant quantum memories $L$ and $R$ which can be atomic ensembles
as in the original DLCZ scheme or single atom trapped in cavity or
some other quantum memories used in \cite{QR6, QR7}. $L$ and $R$ are
both simultaneously illuminated by a pumping laser, and then with a
small probability $L$ or $R$ will jump to another quantum state and
emit one photon to the channel. The path information of the single
photon emitted from $L$ or $R$ will be erased and then this
single-photon enters the detector at the middle point. Now $L$ and
$R$ will be entangled. This heralded entanglement generation is also
a built-in entanglement purification process \cite{QR3}. The lower
the probability of photon emission is, the higher the fidelity of
the resulted entangled-states is. However this is also a
single-photon interference process which is sensitive to the phase
drift due to unknown length drift between the path from the $L$ to
the middle point and the path from $R$ to the middle point. Since
the elementary entanglement generation and entanglement swapping are
all probabilistic, one must keep this phase constant over a very
long time. In fact this demand is beyond modern technology. One can
see Refs. \cite{QR8, QR9} for the detailed analysis of the phase
stability problem of DLCZ.

  To overcome phase instability, Refs. \cite{QR8, QR9} proposed a
new quantum repeaters architecture based on two-photon
Hong-Ou-Mandel-type \cite{Hong-Ou-Mandel} interference. Unlike the
original DLCZ-type quantum repeaters which need to keep the drift of
arms of the long distance interferometer below the wave-length
scale, this new architecture only requires us to keep this drift
below the coherence-length scale of the flying photons, which
greatly facilitates the implementation of the long-distance quantum
repeaters. Another choice is to use the Sagnac interferometer
configuration. With the help of the optical switches, the pumping
pulse to the $L$ should be reflected by a mirror in the $R$ before
the excitation process, and similarly, the pumping pulse to the $R$
should be reflected by an mirror in the $L$ before the excitation
process. With the Sagnac interferometer configuration, the
single-photon from the $L$ and $R$ will have the same path-length if
the length drift of the channel is negligible during the travel time
of the flying photons in the channel. And in the Ref \cite{phase
noise}, it has been proved that one can obtain high enough
interference visibility even the fiber length up to $75km$ with the
Sagnac interferometer configuration.

  However, the above two solutions to the phase instability of
DLCZ-type quantum repeaters still suffer from the polarization
disturbance of the channel. For the two-photon Hong-Ou-Mandel-type
interference quantum repeaters proposed by Refs. \cite{QR8, QR9},
the polarization disturbance of the channel may introduce error in
the elementary entanglement generation step. Suppose the channel
between $L$ and the middle point maps the horizontal polarization
state $|H\rangle$ to the $|F\rangle=(|H\rangle+|V\rangle)/\sqrt{2}$
and the vertical polarization state $|V\rangle$ to the
$|S\rangle=(|H\rangle-|V\rangle)/\sqrt{2}$, then the type-I
Bell-States-Measurement (BSM-I) performed at the middle point cannot
distinguish the error case of both the two ensembles of $L$ emitting
one photon from the correct case. This may be an extreme case, since
the channel disturbance should not introduce such a big change of
polarization. But we can conclude that the polarization disturbance
indeed introduces errors to this quantum repeaters proposal. For the
Sagnac interferometer method, the different polarization
disturbances between the two channels from $L$ and $R$ to the middle
point obviously lower the interference visibility. Of course,
polarization controllers may be used to alleviate the polarization
disturbance, as demonstrated in the experiment of the Ref.
\cite{phase noise}. But for the real-life quantum repeaters the
application of polarization controllers will make the system more
complicated and this method cannot eliminate some polarization
disturbances due to rapid vibrations of the fiber channel.

  Here, in this article we propose a new quantum repeaters architecture
that utilizes the invariance of the anti-symmetric Bell state
$|\psi^-\rangle=(|H\rangle|V\rangle-|V\rangle|H\rangle)/\sqrt{2}$
under collective noise. In fact, when the coherence time of photons
is larger than the delay time resulting from the polarization mode
dispersion, the polarization disturbance can be treated as an
unitary transformation only acting on polarization space \cite{pu1,
pu2}. Our scheme is totally free of polarization disturbance of the
channel. Through Sagnac interferometer configuration, the phase
stability can be also achieved if the distance between neighboring
nodes is not too long. Though topological structure of our scheme is
similar to those in the Refs. \cite{QR8, QR9}, the physical
implementation should not be atomic ensembles, since one ensemble
may emit two or more photons. The single photon source type quantum
memory, such as single atom trapped in cavity may be appropriate for
our scheme. The detailed discussion of our scheme will be given in
section II and a conclusion will be given in section III.

\section{architecture}

 The basic architecture for distributing entanglement for neighboring
nodes of our scheme is shown in Fig. 1.

\begin{figure}[!h]
\center \resizebox{8.5cm}{!}{\includegraphics{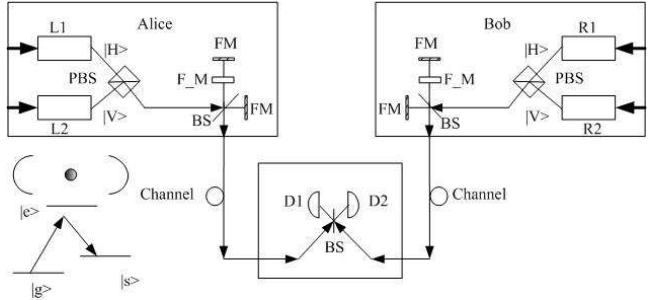}}
\caption{BS: 50:50 beam-splitter; D1 and D2 represent two single
photon detectors respectively; FM: Faraday mirror; F\_M: phase
modulator; PBS: polarization beam-splitter which transmits
horizontal polarized photons and reflects vertical polarized
photons}\label{schematic}
\end{figure}

 As in Fig. 1., Alice and Bob are two neighboring nodes. The
memory units $L_1$, $L_2$ , $R_1$ and $R_2$ should be some
single-photon source type quantum-memories, such as a single
three-level atom trapped in a cavity shown in Fig. 1. Initially, the
four three-level atoms, each in the $L_1$, $L_2$ , $R_1$ and $R_2$
respectively, are all in the ground state $|g\rangle$. We
simultaneously pump the four atoms and with a small probability $p$,
one of these four atoms emits Stokes photon with frequency
$\omega_{es}=\omega_e-\omega_s$. In fact the pumping laser for the
quantum memories $L_1$, $L_2$ must be reflected by a mirror in R
before the pumping process and so does the pumping laser for $R_1$
and $R_2$. For simplicity, we do not draw this Sagnac interferometer
configuration in Fig. 1. We only concern the case that the two
single-photon detectors have two clicks. With probability $p^2$, two
of the four atoms emit one photon respectively, and this could be
given in the operators form:

\begin{equation}
\begin{aligned}
|\Psi\rangle=(&1+\sqrt{p}(S^\dagger_{L_1}H^\dagger_L+S^\dagger_{L_2}V^\dagger_L+S^\dagger_{R_1}H^\dagger_R+S^\dagger_{R_2}V^\dagger_R)\\
&+p(S^\dagger_{L_1}S^\dagger_{L_2}H^\dagger_LV^\dagger_L+S^\dagger_{R_1}S^\dagger_{R_2}H^\dagger_RV^\dagger_R\\
&+S^\dagger_{L_1}S^\dagger_{R_1}H^\dagger_LH^\dagger_R+S^\dagger_{L_2}S^\dagger_{R_2}V^\dagger_LV^\dagger_R\\
&+S^\dagger_{L_1}S^\dagger_{R_2}H^\dagger_LV^\dagger_R+S^\dagger_{L_2}S^\dagger_{R_1}V^\dagger_LH^\dagger_R)+O(p))|g\rangle_{LR}|0\rangle_p
\end{aligned}
\end{equation}
in which, $S^\dagger=|s\rangle\langle g|$ means the atomic
transition from the ground level $|g\rangle$ to the meta-stable
level $|s\rangle$, meanwhile $H^\dagger$ and $V^\dagger$ represent
the creation operators for the horizontal and vertical polarized
photons respectively. Before the photons enter the channel, they
will pass through the unbalanced interferometer consisting of two
Faraday mirrors and one polarization-depended phase modulator in the
long arm of the interferometer. The function of Faraday mirrors is
to eliminate the possible polarization disturbance of the
interferometer \cite{F-M1, F-M2}. The phase modulator in the long
arm of the unbalanced interferometer only adds $\pi$ phase to the
horizontal polarized photons and does not change the vertical
polarized photons. Only concerning the events which may introduce
two clicks to the detector 1 and detector 2 and assuming the
beam-splitter will add $\pi/2$ phase to the reflected photons and
won't change the transmitting photons, we can deduce that:

\begin{equation}
\begin{aligned}
&H^\dagger_LV^\dagger_L\rightarrow
\frac{1}{\sqrt{2}}(\frac{1}{\sqrt{2}}(H^\dagger_{1L}V^\dagger_{1L}-H^\dagger_{2L}V^\dagger_{2L})
+\frac{1}{\sqrt{2}}(H^\dagger_{1L}V^\dagger_{2L}-V^\dagger_{1L}H^\dagger_{2L}))\\
&H^\dagger_RV^\dagger_R\rightarrow
\frac{1}{\sqrt{2}}(\frac{1}{\sqrt{2}}(H^\dagger_{1R}V^\dagger_{1R}-H^\dagger_{2R}V^\dagger_{2R})
+\frac{1}{\sqrt{2}}(H^\dagger_{1R}V^\dagger_{2R}-V^\dagger_{1R}H^\dagger_{2R}))\\
\end{aligned}
\end{equation}
in which, the subscript $1(2)$ represents the time-bin of the
photons passing through the long arm and short arm of interferometer
respectively, which result in the two different time-bins, and
subscript $L(R)$ represents the left(right) channel as in Fig. 1.
The second term of the right side of the equation (2) is just the
anti-symmetric $|\psi^-\rangle$ state. From this equation we know if
the two photons occupy the two different time-bins, they will be in
the $|\psi^-\rangle$ state. Conversely, if the two photons occupy
the same time-bin 1 or 2, it will not be $|\psi^-\rangle$ state.
According to Ref. \cite{pu1}, polarization disturbance can be well
approximated by a collective unitary transformation as long as the
delay between the two time-bins is small compared to the variation
of disturbance, which has been verified by experiment \cite{pu3}.
Therefore through the invariance of $|\psi^-\rangle$ state and
erasing the path information, the entanglement can be well
established. Now we can conclude that if the detector 1 and detector
2 get one click in each of the two time-bins, we can get the
entangled-states given by

\begin{equation}
\begin{aligned}
&S^\dagger_{L_1}S^\dagger_{L_2}H^\dagger_LV^\dagger_L+S^\dagger_{R_1}S^\dagger_{R_2}H^\dagger_RV^\dagger_R
\rightarrow\\
&\frac{1}{2\sqrt{2}}(S^\dagger_{L_1}S^\dagger_{L_2}-S^\dagger_{R_1}S^\dagger_{R_2})(-H^\dagger_{1D_1}V^\dagger_{2D_1}+V^\dagger_{1D_1}H^\dagger_{2D_1}
+H^\dagger_{1D_2}V^\dagger_{2D_2}\\&-V^\dagger_{1D_2}H^\dagger_{2D_2})\\
&+i\frac{1}{2\sqrt{2}}(S^\dagger_{L_1}S^\dagger_{L_2}+S^\dagger_{R_1}S^\dagger_{R_2})(H^\dagger_{1D_1}V^\dagger_{2D_2}+H^\dagger_{1D_2}V^\dagger_{2D_1}
-V^\dagger_{1D_1}H^\dagger_{2D_2}\\&-V^\dagger_{1D_2}H^\dagger_{2D_1})\\
\end{aligned}
\end{equation}
If we get two clicks in the same detector we obtain
$(S^\dagger_{L_1}S^\dagger_{L_2}-S^\dagger_{R_1}S^\dagger_{R_2})/\sqrt{2}$,
and if one click in each detector
$(S^\dagger_{L_1}S^\dagger_{L_2}+S^\dagger_{R_1}S^\dagger_{R_2})/\sqrt{2}$
is prepared.

 Meanwhile, events described by $S^\dagger_{L_1}S^\dagger_{R_1}H^\dagger_LH^\dagger_R$,
$S^\dagger_{L_2}S^\dagger_{R_2}V^\dagger_LV^\dagger_R$,
$S^\dagger_{L_1}S^\dagger_{R_2}H^\dagger_LV^\dagger_R$,
$S^\dagger_{L_2}S^\dagger_{R_1}V^\dagger_LH^\dagger_R$ in the
equation (1) are also probable to click the detectors twice, e.g.
both $L_1$ and $R_1$ both emit horizontal polarized photons.
Obviously, photons as described by  $H^\dagger_LH^\dagger_R$,
$V^\dagger_LV^\dagger_R$, $H^\dagger_LV^\dagger_R$ or
$V^\dagger_LH^\dagger_R$ cannot constitute the $|\psi^-\rangle$
state. Hence these photon-states suffer from random polarization
disturbance, and we cannot predict the states of these photons after
traveling along the two different channels. Therefore, in this
elementary entanglement distributing step, we cannot eliminate these
errors. Assuming we get the first click in detector 1 and the second
click in detector 2 and neglect dark counts of the detectors, what
we get will be a mixture given by:

\begin{equation}
\begin{aligned}
\rho_{LR}=&\frac{1}{\sqrt{6}}(\sqrt{2}\rho^{+LR}_{ENG}+|sg\rangle_{L_1L_2}\langle
sg|\otimes |sg\rangle_{R_1R_2}\langle
sg|\\&+|gs\rangle_{L_1L_2}\langle gs|\otimes
|gs\rangle_{R_1R_2}\langle gs|+|sg\rangle_{L_1L_2}\langle sg|\otimes
|gs\rangle_{R_1R_2}\langle gs|\\&+|gs\rangle_{L_1L_2}\langle
gs|\otimes |sg\rangle_{R_1R_2}\langle sg|)
\end{aligned}
\end{equation}
in which, $\rho^{+LR}_{ENG}=|\Psi^+\rangle_{LR}\langle\Psi^+|$ and
$|\Psi^+\rangle_{LR}=(S^\dagger_{L_1}S^\dagger_{L_2}+S^\dagger_{R_1}S^\dagger_{R_2})|g\rangle_{L_1L_2R_1R_2}/\sqrt{2}$.
Note that in the above equation we have assumed that all error
events do not result in any entanglement between $L$ and $R$.
Obviously this assumption is just for simplicity and does not spoil
the preciseness in this paper, since all error events can be
eliminated by the entanglement swapping step. The first item of the
above equation represents the correct case while the others
represent the error cases. From the above equation, we know there
are many errors after the first entanglement distributing step.
Fortunately, all these possible errors can be eliminated with a
simple entanglement swapping step, which is shown in Fig. 2.

\begin{figure}[!h]
\center \resizebox{8.5cm}{!}{\includegraphics{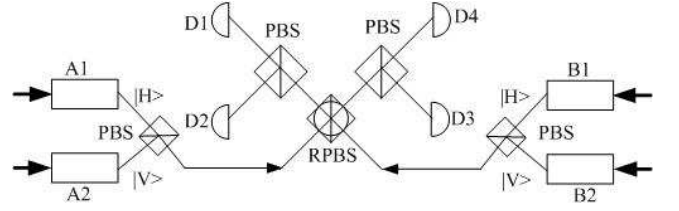}}
\caption{D1, D2, D3 and D4 represent four photon-number-resolvable
detectors respectively; PBS: polarization beam-splitter which
transmits horizontal polarized photons $|H\rangle$ and reflects
vertical polarized photons $|V\rangle$; RPBS: rotated polarization
beam-splitter which transmits photons with polarization
$|F\rangle=(|H\rangle+|V\rangle)/\sqrt{2}$ and reflects photons
photons with polarization
$|S\rangle=(|H\rangle-|V\rangle)/\sqrt{2}$}\label{schematic}
\end{figure}

 Suppose that we have successfully performed entanglement
distributing between nodes $L$, $A$ and $B$, $R$, which means that
the $L$, $A$ and $B$, $R$ are in the mixture $\rho_{LA}$ and
$\rho_{BR}$ respectively. The entanglement swapping process is same
as those in Refs. \cite{QR3, QR8, QR9}, in which we pump the atoms
in $A$ and $B$ to excite the transition from $|s\rangle$ to
$|g\rangle$. From the equation (4), the correct case that the $L$,
$A$ are in the state
$|\Psi^+\rangle_{LA}=(S^\dagger_{L_1}S^\dagger_{L_2}+S^\dagger_{A_1}S^\dagger_{A_2})|g\rangle_{L_1L_2A_1A_2}/\sqrt{2}$
and the $B$, $R$ are in the state
$|\Psi^+\rangle_{BR}=(S^\dagger_{B_1}S^\dagger_{B_2}+S^\dagger_{R_1}S^\dagger_{R_2})|g\rangle_{B_1B_2R_1R_2}/\sqrt{2}$
, has a probability of $1/9$. Note that the $A$ and $B$ are in the
same location and thus any polarization or phase noise is
negligible. For this correct case and while only components that
will induce detectors double clicks are considered, we deduce that
the successful entanglement swapping process is given by:

\begin{equation}
\begin{aligned}
|\Psi^+\rangle_{LA}|\Psi^+\rangle_{BR}\rightarrow&
\frac{1}{8}(\frac{2\sqrt{2}}{\sqrt{2}}(S^\dagger_{L_1}S^\dagger_{L_2}+S^\dagger_{R_1}S^\dagger_{R_2})(H^\dagger_1V^\dagger_2+H^\dagger_4V^\dagger_3)\\&
+\frac{\sqrt{2}}{\sqrt{2}}(S^\dagger_{L_1}S^\dagger_{L_2}-S^\dagger_{R_1}S^\dagger_{R_2})(H^{\dagger
2}_1+V^{\dagger 2}_2-H^{\dagger 2}_4-V^{\dagger 2}_3))\\&
|g\rangle_{LR}|0\rangle_p
\end{aligned}
\end{equation}
in which the subscript $1$, $2$, $3$ and $4$ in the photon creations
operators represent the corresponding modes for the photon
detectors. According to the above equation, if we have two clicks in
detector 1 and 2 or 3 and 4, the entanglement swapping is successful
and thus entanglement between $L$ and $R$ is established with
probability of $1/2$. Here we abandon the $H^{\dagger 2}_1$,
$V^{\dagger 2}_2$, $H^{\dagger 2}_4$ and $V^{\dagger 2}_3$, since
these events may be introduced by error cases in equation (4), which
will be discussed in the followings.

 Concretely speaking, if one of $L$, $A$ and $B$, $R$ is in the correct case and
the other is in the wrong case, obviously the detectors will have
only 1 click or 3 clicks. Thus we can distinguish this case from the
correct case. If both the $L$, $A$ and $B$, $R$ are in the wrong
case, the entanglement swapping process will result in the following
photons creations given by:

\begin{equation}
\begin{aligned}
H^\dagger_AH^\dagger_B\rightarrow
&(H^\dagger_4+V^\dagger_3)(H^\dagger_1+V^\dagger_2)+(H^\dagger_4+V^\dagger_3)(H^\dagger_1+V^\dagger_2)\\&+H^{\dagger
2}_1-V^{\dagger 2}_2+H^{\dagger 2}_4-V^{\dagger 2}_3\\
V^\dagger_AV^\dagger_B\rightarrow
&(H^\dagger_4+V^\dagger_3)(H^\dagger_1+V^\dagger_2)+(H^\dagger_4+V^\dagger_3)(H^\dagger_1+V^\dagger_2)\\&-H^{\dagger
2}_1+V^{\dagger 2}_2-H^{\dagger 2}_4+V^{\dagger 2}_3\\
H^\dagger_AV^\dagger_B\rightarrow
&(H^\dagger_4+V^\dagger_3)(H^\dagger_1+V^\dagger_2)-(H^\dagger_4+V^\dagger_3)(H^\dagger_1+V^\dagger_2)\\&+H^{\dagger
2}_1-V^{\dagger 2}_2-H^{\dagger 2}_4+V^{\dagger 2}_3\\
V^\dagger_AH^\dagger_B\rightarrow
&(H^\dagger_4+V^\dagger_3)(H^\dagger_1+V^\dagger_2)-(H^\dagger_4+V^\dagger_3)(H^\dagger_1+V^\dagger_2)\\&-H^{\dagger
2}_1+V^{\dagger 2}_2+H^{\dagger 2}_4-V^{\dagger 2}_3
\end{aligned}
\end{equation}
According to equation (6), obviously neither of these items results
in $H^\dagger_1V^\dagger_2$ and $H^\dagger_4V^\dagger_3$. Therefore,
the two clicks in both detectors 1 and 2 or  3 and 4 will project
the $L$ and $R$ into the desired entangled-state
$|\Psi^+\rangle_{LR}$, while all the error cases cannot be projected
into this state.

 Now we can conclude that: as in Fig.2., when neighboring nodes $L$, $A$ and
$B$, $R$ are entangled respectively, the probability of successful
entanglement swapping is $\frac{1}{9}\times
\frac{1}{4}=\frac{1}{36}$, which is much smaller than $\frac{1}{2}$
as in original DLCZ scheme \cite{QR3}. Although the decrease of
efficiency of entanglement swapping will lower the total efficiency,
the scalability of our scheme is not affected, since the decrease of
efficiency can be regarded as the inefficiency of the detectors,
which is same as DLCZ scheme. We also note that in further
entanglement swapping steps, this efficiency is $\frac{1}{2}$, just
same as DLCZ scheme, since all errors have been eliminated in the
elementary entanglement swapping step.

 The further applications of the final entangled state $|\Psi^+\rangle_{LR}$,
such as quantum cryptography, is same as original DLCZ scheme, since
the final entangled state $|\Psi^+\rangle_{LR}$ is equivalent to the
one in DLCZ. Obviously with two pairs of such entangled states, two
phase modulators and several optical switches, a phase-modulation
type quantum key distribution can be implemented easily. One can see
Ref \cite{QR3} for detailed.

\section{conclusion}

 The original DLCZ-type heralded quantum repeaters suffer from the
phase noise and polarization disturbance of the channel. The Sagnac
interferometer configuration may overcome the phase noise when the
drift of the channel length is not too fast, but still needs
feedback polarization controllers. The two-photon Hong-Ou-Mandel
interference type quantum repeaters may be free of phase noise but
also vulnerable to the polarization disturbance. Thus neither of
these solutions can overcome the polarization disturbance.

 In this paper, we propose a new quantum repeaters architecture,
which is based on the invariance of anti-symmetric Bell state
$|\psi^-\rangle=(|H\rangle|V\rangle-|V\rangle|H\rangle)/\sqrt{2}$
under collective unitary transformation. This new setup can be
totally free of polarization disturbance of channel. With Sagnac
interferometer configuration, phase noise can be also alleviated.
Single photon source quantum memory should be adopted for physical
implementation of this proposal, e.g. single three-level atom
trapped in high-finesse cavity is a good candidate. With an
unbalanced Faraday-Michelson interferometer we can project the
two-photon $|H\rangle|V\rangle$ wave-packet into the
$|\psi^-\rangle$ state distinguishable by two very near time-bins.
Although the elementary entanglement distributing may introduce some
errors, the following entanglement swapping step can eliminate all
possible errors. In a word, the final entanglement state can be of
high fidelity and immune to both polarization and phase disturbance
of the channels.

 This work was supported by National Fundamental Research Program
of China (2006CB921900), National Natural Science Foundation of
China (60537020, 60621064) and the Innovation Funds of Chinese
Academy of Sciences.
*To whom correspondence should be addressed,
Email: zfhan@ustc.edu.cn.

\end{document}